\documentclass[3p,times,twocolumn]{elsarticle}
 \biboptions{comma,sort&compress}
 
\usepackage{graphicx}
\usepackage{here}
\usepackage{ecrc}

\volume{00}

\firstpage{1}

\journalname{Nuclear and Particle Physics Proceedings}

\runauth{}

\jid{nppp}

\jnltitlelogo{Nuclear and Particle Physics Proceedings}

\usepackage{amssymb}

\usepackage{lineno}

\usepackage[figuresright]{rotating}

\begin{document}

\begin{frontmatter}

\title{ Measurement of quarkonium production in ALICE
 $^*$}
 \cortext[cor0]{Talk given at 23rd International Conference in Quantum Chromodynamics (QCD 20),  27 - 30 October 2020, Montpellier - FR}
 \author{Victor Feuillard, for the ALICE Collaboration}
\ead{victor.jose.gaston.feuillard@cern.ch}
\address{Physikalisches Institut, Heidelberg, Germany}

\pagestyle{myheadings}
\markright{ }
\begin{abstract}
ALICE is designated to study the quark--gluon plasma (QGP), a state of matter where, due to high temperature and density, quarks and gluons are deconfined. One of the probes used to investigate this state of matter is quarkonium states, bound states of either a charm and anti-charm quark pair (charmonia) or a bottom and anti-bottom quark pair (bottomonia). The presence of the QGP is expected to modify the quarkonium production yields in a very specific way due to a balance between medium-induced suppression, and a recombination mechanism or a hadronization mechanism. To understand the the properties of the QGP in nucleus-nucleus collisions, it is essential to measure the quarkonium differential yields in proton--proton collisions, as it provides a reference and allows the investigation of quarkonium production mechanisms, as well as in proton--nucleus collisions to understand the cold nuclear matter effects that appear. In this contribution, the latest results for quarkonium production measured with the ALICE detector in pp collisions at different collision energies are reported. The measurements of the nuclear modification factor and anisotropic flow in Pb--Pb collisions at $\sqrt s\textsubscript{NN} = 5.02$~TeV and in p--Pb at $\sqrt s\textsubscript{NN} =8 .16$~TeV at mid- and forward rapidity are also reported. All measurements are compared to various theoretical predictions.
\end{abstract}
\begin{keyword}  
QGP \sep heavy-ion \sep quarkonium \sep J/$\psi$ \sep $\Upsilon$ \sep elliptic flow \sep Cold Nuclear Matter \sep ALICE \sep LHC

\end{keyword}

\end{frontmatter}
\section{Introduction}

The quark--gluon plasma (QGP) is a state of matter theoretically predicted by lattice quantum chromodynamics (QCD) where quark and gluons are deconfined. There is a particular interest in studying the QGP since in current cosmological models it was the first state of matter in the early stages of the Universe (up to $\tau\approx10$~$\mu$s). Experimentally it is possible to create a QGP using ultra-relativistic heavy-ion collisions, like at RHIC~\cite{ROSER200223} or the LHC~\cite{Evans:2008zzb}, but only within a short period of time and a very small volume ($\sim$10~fm/\textit{c} and $\sim$10\textsuperscript{3}~fm\textsuperscript{3} in Pb--Pb at $\sqrt s\textsubscript{NN} = 2.76$~TeV)~\cite{Aamodt:2011mr}. Due to this small time period and volume, it is impossible to observe the QGP directly, therefore measurements of the particles produced, such as quarkonia, are used as probes.
Because of their large mass, charm and bottom quarks are primarily produced at the very beginning of the collision and experience the entire medium evolution. Quarkonium resonances, which are bound states of a $\rm{c\bar{c}}$ pair for charmonium or a $\rm{b\bar{b}}$ pair for bottomonium, are therefore among the most direct signatures for the QGP formation.
Firstly, theory predicts that quarkonia are suppressed in a QGP due to the colour screening: the presence of free colour charges in the medium, the binding potential is screened~\cite{Matsui:1986dk}. This leads to a reduction of the number of quarkonium states produced. 
Secondly, a competing mechanism can occur, namely recombination: if there are enough quark pairs produced, and they thermalize in the QGP, then quarkonia can be regenerated by the recombination of these quark pairs, resulting in an increase of the quarkonium yields \cite{BraunMunzinger:2000px,Thews:2000rj}. 

In the following, a selection of the latest results obtained by ALICE in pp, p--Pb, and Pb--Pb collisions are reported. A complete description of the ALICE detector can be found in~\cite{Aamodt:2008zz,Abelev:2014ffa}.

\section{Results in pp collisions at $\sqrt{s} = 13$~TeV}
In addition to the measurements in heavy-ion collisions, it is also interesting to measure the charmonium production in pp collisions. Indeed, it provides a reference for the measurement in Pb--Pb collisions and also allows to explore the QCD models, as the quarkonium formation involves both hard-scale processes for heavy quark production, and soft-scale processes for hadronization.

\subsection{J/$\psi$ production cross section}
\begin{figure}[htb]
\begin{center}
\vspace{9pt}
\includegraphics[scale=0.32]{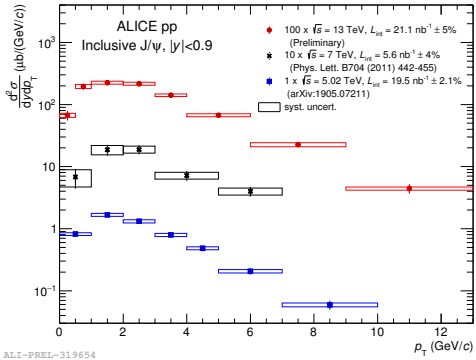}
\caption{J/$\psi$ cross section measured at mid-rapidity for different collision energies as a function of $p_{\rm{T}}$.}
\label{fig:JPsi_CrossSection}
\end{center}
\end{figure}

A new measurement of the J/$\psi$ production cross section in pp collisions at $\sqrt{s} = 13$~TeV at mid-rapidity as a function of transverse momentum ($p\textsubscript{T}$) is presented in Fig.~\ref{fig:JPsi_CrossSection}. It is compared with previous measurements at mid-rapidity at collision energies of $\sqrt{s} = 7$~TeV~\cite{Aamodt:2011gj} and $\sqrt{s} = 5.02$~TeV~\cite{Acharya:2019lkw}. With the increasing collision energy, a hardening of the $p\textsubscript{T}$ is observed, similar to observations at forward rapidity~\cite{Acharya:2017hjh}.

\subsection{Normalized J/$\psi$ yield as a function of charged-particle multiplicity}
A good test of QCD models is the measurement of the normalized J/$\psi$ yield as a function of the normalized charged-particle multiplicity. The correlation between the two quantities helps to constrain the interplay between the soft and hard mechanisms in pp collisions. 

\begin{figure}[htb]
\begin{center}
\vspace{9pt}
\includegraphics[scale=0.32]{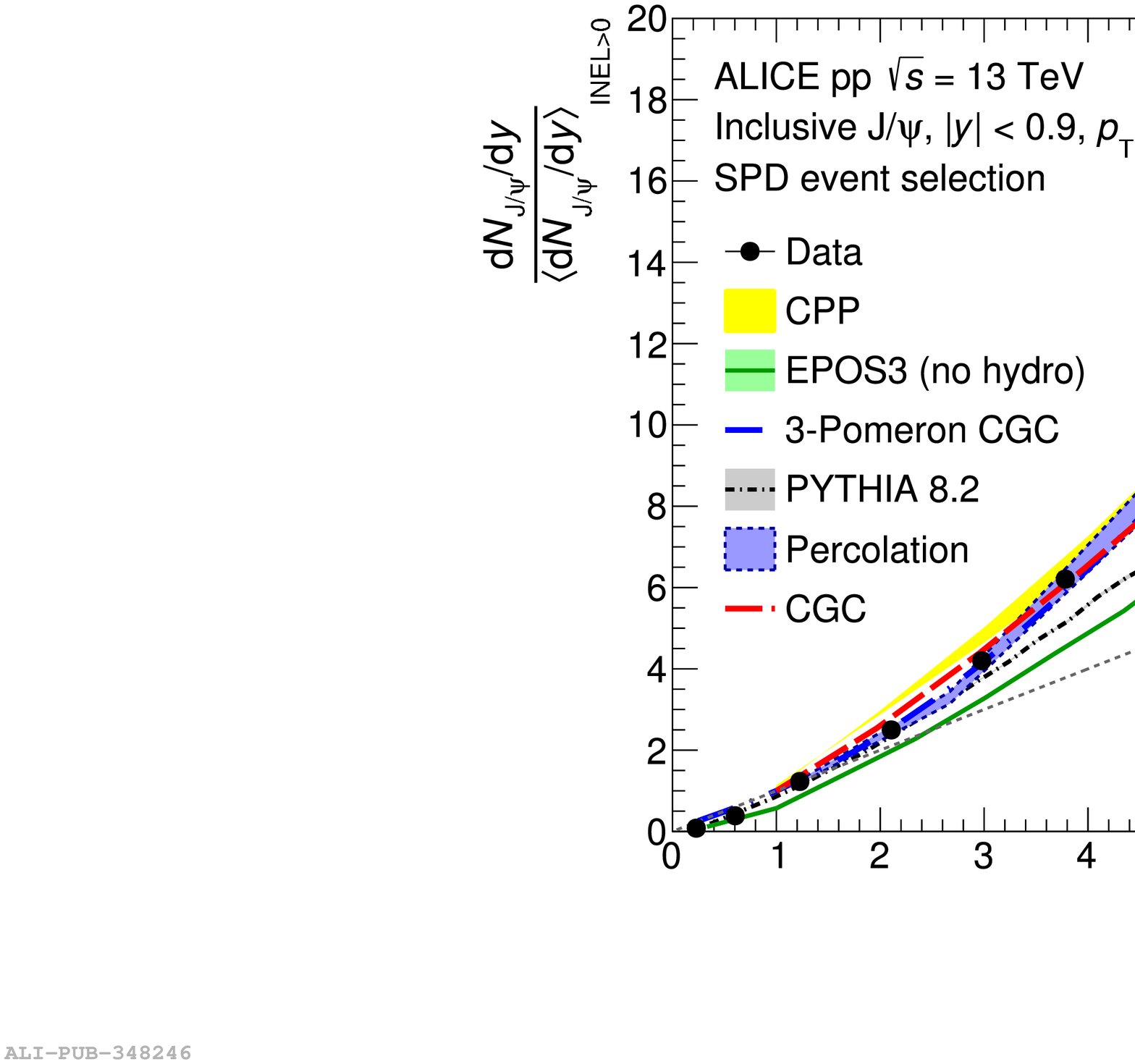}
\caption{Normalized inclusive $p_{\rm{T}}$-integrated J/$\psi$ yield at mid-rapidity as a function of normalized charged-particle pseudorapidity density ($|\eta|<1$)~\cite{Acharya:2020pit} compared with different models~\cite{Kopeliovich:2013yfa,Ferreiro:2012fb,Siddikov:2019xvf,Ma:2018bax,Werner:2013tya}.}
\label{fig:NormJPsi_vs_Mult}
\end{center}
\end{figure}

A new measurement of the normalized inclusive J/$\psi$ yield at mid-rapidity as a function of normalized charged-particle pseudorapidity density at mid-rapidity in pp collisions at $\sqrt{s} = 13$~TeV is presented in Fig.~\ref{fig:NormJPsi_vs_Mult}. The data are compared with several theoretical models~\cite{Acharya:2020pit}. The normalized J/$\psi$ yield exhibits a faster than linear increase with the normalized multiplicity. This behaviour is predicted by all the theoretical models currently used. It is explained  effectively as the result of a reduction of the charged-particle multiplicity at high multiplicity, however each model attributes the observed behavior to different underlying processes (color string reconnection or percolation, gluon saturation, coherent particle production, 3-gluon fusion in gluon ladders/Pomerons). 

In order to better understand the mechanisms at play, more tests are required on the models. In addition, the possibility to separate between prompt and non-prompt J/$\psi$ could provide further information.

\section{Results in p--Pb collisions at $\sqrt{s\textsubscript{NN}} = 8.16$~TeV}
Measuring the nuclear modification in p--Pb collisions, meaning the ratio of production yields in p--Pb collisions with respect to pp collisions, allows to understand cold nuclear matter (CNM) effects, such as nuclear shadowing of the partonic structure functions. It leads to a change in the probability for a quark or gluon to carry a fraction of the nucleon momentum ($x$) and therefore affects the production cross section of the heavy quark pair. This allows to distinguish in the Pb--Pb measurements the effects that originate from the hot medium from those caused by the presence of nuclear matter. The asymmetry of the p--Pb collision allows us to probe different $x$ regions of the lead nucleus: measuring in the p-going direction equates to investigating the low-$x$, whereas  measuring in the Pb-going direction corresponds to investigating the high-$x$ region.

\subsection{Charmonium nuclear modification factor}

\begin{figure}[htb]
\begin{center}
\vspace{9pt}
\includegraphics[scale=0.34]{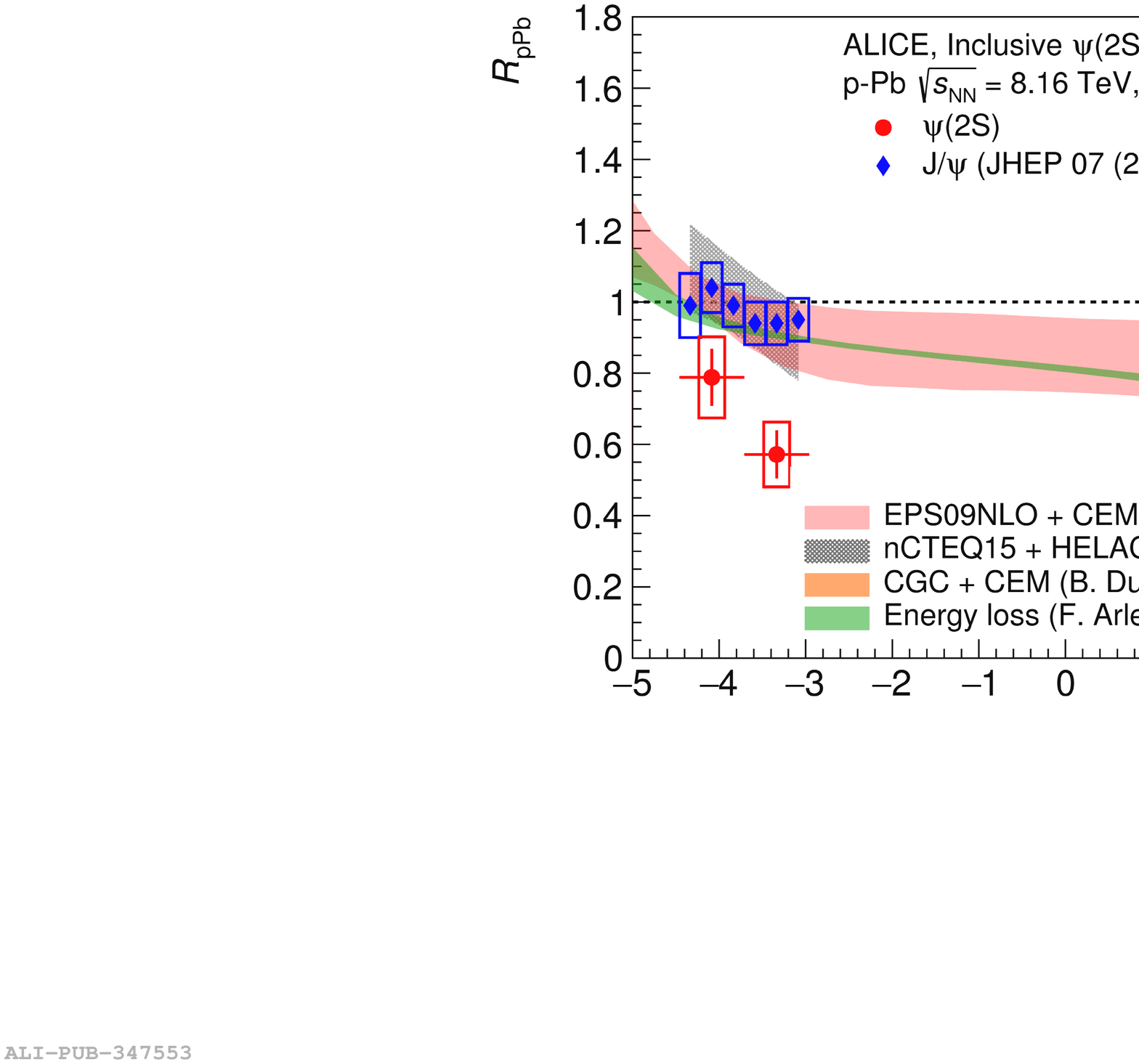}
\includegraphics[scale=0.34]{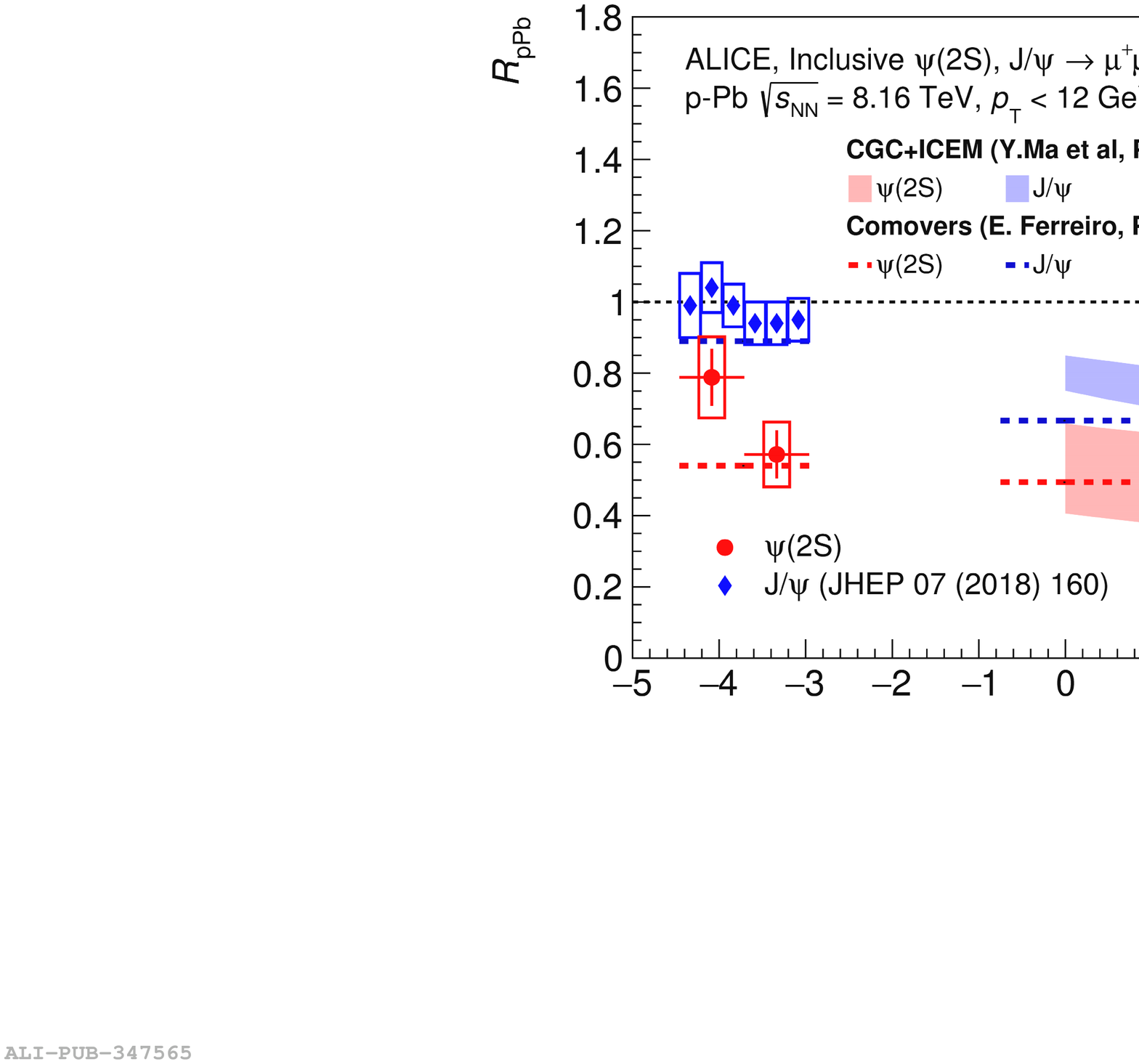}
\caption{(Top) J/$\psi$ and $\psi$(2S) $R\textsubscript{pPb}$ as a function of rapidity compared with models including CNM only~ \cite{Acharya:2020wwy}. (Bottom) Same data compared with models including final state effects.}
\label{fig:JPsi_RpPb}
\end{center}
\end{figure}

A new measurement of the $\psi$(2S) nuclear modification factor ($R\textsubscript{pPb}$) in p--Pb collisions at $\sqrt{s\textsubscript{NN}} = 8.16$~TeV as a function of rapidity is presented in Fig.~\ref{fig:JPsi_RpPb}, compared with the J/$\psi$ one~\cite{Acharya:2020wwy}. We can observe that at forward rapidity, corresponding to the p-going direction, the J/$\psi$ and the $\psi$(2S) are suppressed in p--Pb collisions. The values for the J/$\psi$ and the $\psi$(2S) agree within uncertainties. On the other hand, at backward rapidity, corresponding to the Pb-going direction, the $\psi$(2S) is also suppressed whereas the J/$\psi$ $R$\textsubscript{pPb} is compatible with unity.

In the top panel of Fig.~\ref{fig:JPsi_RpPb}, the data are compared with theoretical models that include CNM effects only~\cite{Ducloue:2016pqr,Albacete:2017qng,Kusina:2017gkz,Arleo:2014oha}, which are largely independent from the specific charmonium resonance and can therefore be compared to both. The models show a good agreement with the J/$\psi$ values both at forward and backward rapidity. However in the case of the $\psi$(2S), the models show a good agreement with the result at forward rapidity but overestimate the result at backward rapidity.

In the bottom panel of Fig.~\ref{fig:JPsi_RpPb}, measurements are compared with theoretical models that include some final state effects as well. These models show a good agreement with both resonance measurements, at both forward and backward rapidity, when available. This tends to indicate that final state interactions, which can lead to a suppression of the charmonium resonance, may have a stronger effect on the $\psi(2S)$ due to its lower binding energy.

\subsection{Bottomonium nuclear modification factor}
\begin{figure}[htb]
\begin{center}
\vspace{9pt}
\includegraphics[scale=0.34]{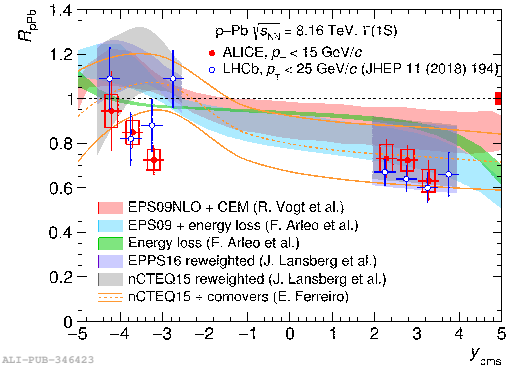}
\caption{$\Upsilon$(1S) $R_{\rm{pPb}}$ as a function of rapidity compared with models and LHCb results~ \cite{Acharya:2019lqc}.}
\label{fig:Upsilon_RpPb}
\end{center}
\end{figure}

In Fig.~\ref{fig:Upsilon_RpPb} the $\Upsilon$(1S) $R\textsubscript{pPb}$ in p--Pb collisions at $\sqrt{s\textsubscript{NN}} = 8.16$~TeV is presented as a function of rapidity~\cite{Acharya:2019lqc}. It is compared with the LHCb results~\cite{Aaij:2018scz} and with several theoretical models~\cite{Albacete:2017qng,Kusina:2017gkz,Arleo:2014oha,Lansberg:2016deg,Ferreiro:2018vmr}. 

The results show a good agreement with the LHCb measurement in the two directions. A suppression of the $\Upsilon$(1S), both at forward and backward rapidity, with a stronger suppression at forward rapidity is seen. When compared with the theoretical models, an overall agreement is observed considering the still significant experimental uncertainties.

\section{Results in Pb--Pb collisions at $\sqrt{s\textsubscript{NN}} = 5.02$~TeV}
\subsection{J/$\psi$ nuclear modification factor}
\begin{figure}[htb]
\begin{center}
\vspace{9pt}
\includegraphics[scale=0.34]{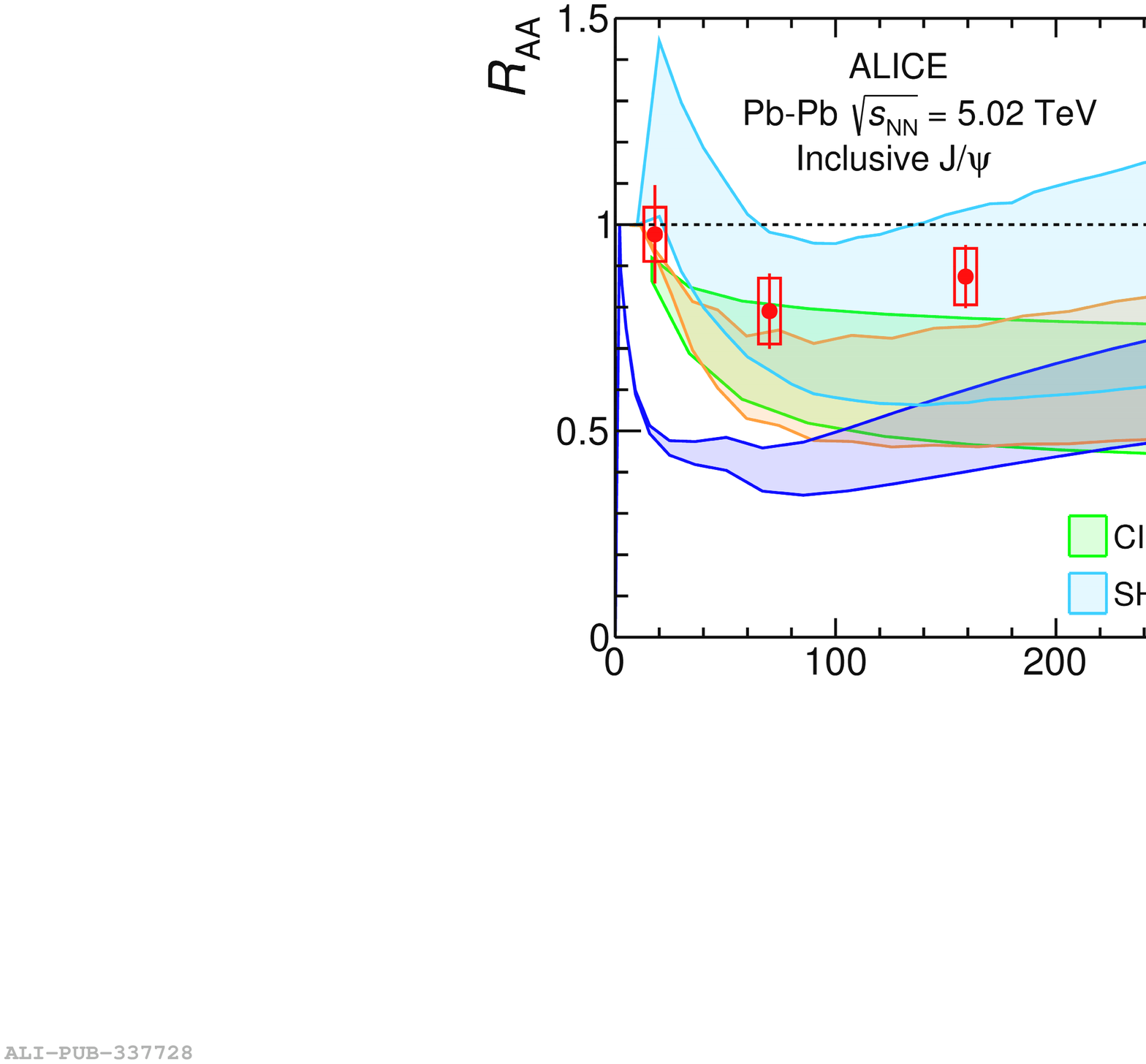}
\caption{$R$\textsubscript{AA} of the J/$\psi$ at $\sqrt{s\textsubscript{NN}} =5.02$~TeV~\cite{Acharya:2019lkh} compared with theoretical models~\cite{Andronic:2007bi,ZHAO2011114,PhysRevC.89.054911,Ferreiro201457}.}
\label{fig:RAAJPsi_Mid}
\end{center}
\end{figure}

The J/$\psi$ nuclear modification factor in Pb--Pb collisions at $\sqrt{s\textsubscript{NN}} = 5.02$~TeV, measured at mid-rapidity is presented in Fig.~\ref{fig:RAAJPsi_Mid} as a function of the mean number of participants~\cite{Acharya:2019lkh}. A moderate suppression for mid-central collisions, and an increase towards central collisions is seen. The measurement in the most peripheral region is compatible with unity. These results strengthen the hypothesis of regeneration, as a model with the color screening alone could not explain the large values nor the increase seen in central collisions.

The result is compared with several theoretical models. The Statistical Hadronization Model (SHM)~\cite{Andronic:2007bi} shows a good agreement with the data, with the uncertainties entirely due to the c$\rm{\bar{c}}$ cross section. The two Transport Models (TM)~\cite{ZHAO2011114,PhysRevC.89.054911} and the Comover Interaction Model (CIM)~\cite{Ferreiro201457} tend to underestimate the measurement in the most central collisions.

When observed as a function of rapidity, together with the measurement made in the dimuon decay channel at forward rapidity, the J/$\psi$ $R\textsubscript{AA}$ exhibits a maximum at mid-rapidity ($R_{\rm{AA}}^{y<|0.9|} = 0.97\pm0.05~\rm{(stat.)}\pm0.1~\rm{(syst.)}$~\cite{Acharya:2019lkh}) and decreases towards large rapidities ($R_{\rm{AA}}^{2.5<y<4.0} = 0.65\pm0.01~\rm{(stat.)}\pm0.05~\rm{(syst.)}$~\cite{Acharya:2019iur}). This result was predicted for the regeneration scenario, due to the larger value of the production cross section $\sigma_{\rm{c\bar{c}}}$ at mid-rapidity, leading to more quark pairs created at the beginning of the collisions, and therefore, more recombination of charmonium. For color screening, the opposite effect is predicted.

\subsection{Anisotropic flow}
The azimuthal dependence of the particle production, the anisotropic flow, is a particularly interesting measurement. It can be expressed as a Fourier decomposition: $\frac{\rm{d}N}{\rm{d}\varphi} \propto 1 + 2 \sum_{\rm{n}} v_{\rm{n}} \cos[\rm{n}(\varphi - \Psi_{\rm{n}})]$, where $\varphi$ is the azimuthal angle, $\Psi_{\rm{n}}$ is the initial state symmetry plane for the n-th harmonic and $v_{\rm{n}}$ is the n-th order Fourier spatial coefficient. The initial anisotropy of the collision is transformed into a momentum anisotropy of the final state particles, that can be quantified by the Fourier coefficients $v\textsubscript{n}$. In particular the $v_2$ coefficient, called the elliptic flow, is caused by the ellipsoidal shape of the overlap region in non-central collisions, and the $v_3$, called the triangular flow, is understood to arise from fluctuations in the initial energy-density profile.

\begin{figure}[htb]
\begin{center}
\vspace{9pt}
\includegraphics[scale=0.395]{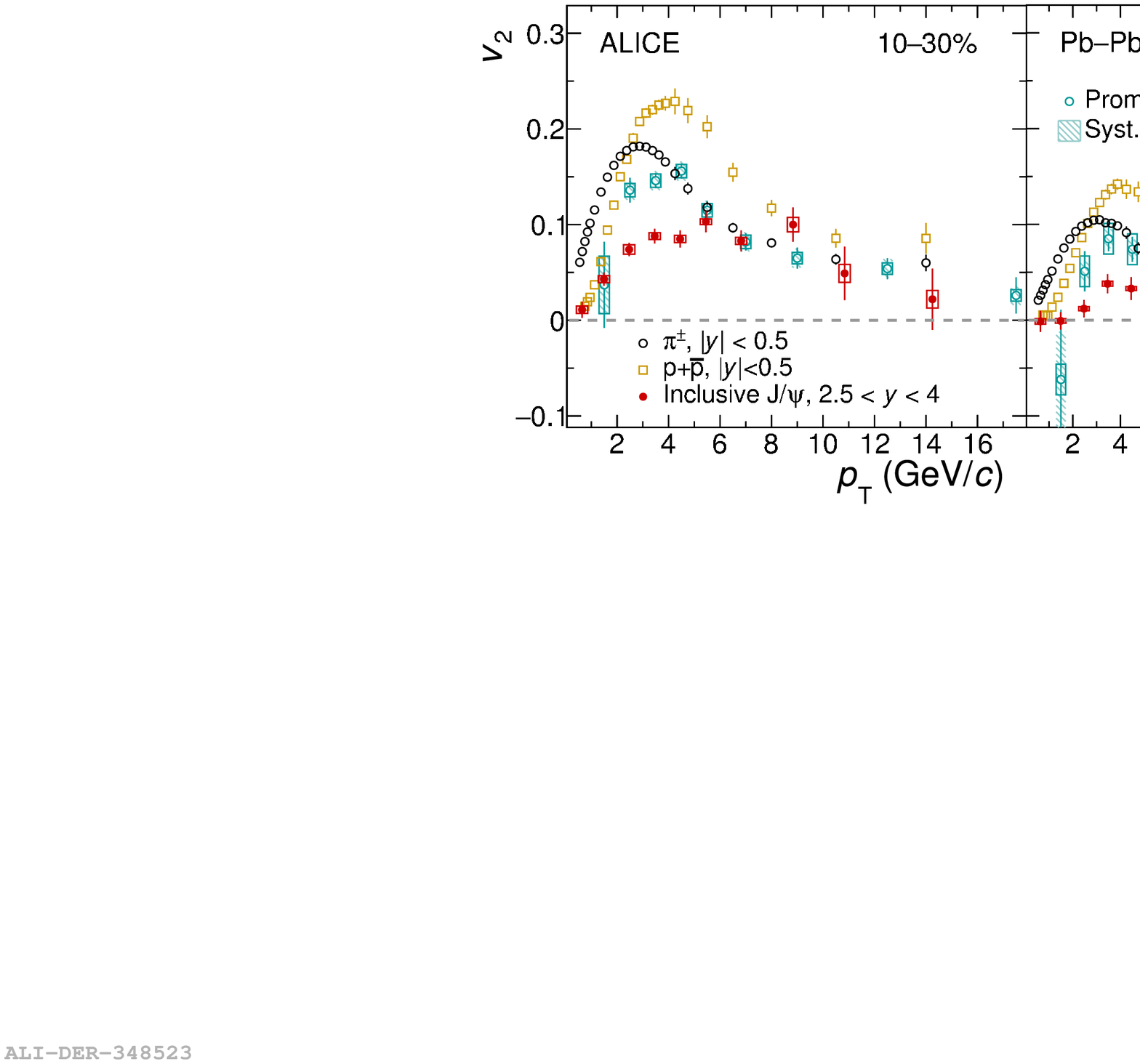}
\caption{J/$\psi$ elliptic and triangular flow as a function of $p_{\rm{T}}$ compared with other species.}
\label{fig:JPsi_v2_v3}
\end{center}
\end{figure}

In Fig.~\ref{fig:JPsi_v2_v3} the elliptic flow of the J/$\psi$ at forward rapidity is presented in the left panel, and the first measurement of the J/$\psi$ triangular flow in ALICE in the right panel, as a function of $p_\textsubscript{T}$~\cite{Acharya:2020jil}. It is compared with the mid-rapidity $v_2$ and $v_3$ of pions, D mesons and protons~\cite{Acharya:2020pnh,Acharya:2018zuq}. The J/$\psi$ elliptic flows is positive and reaches a maximum at intermediate $p_\textsubscript{T}$, before decreasing. In the low $p_\textsubscript{T}$ region ($p_\textsubscript{T} < 6$~GeV/$c$), a mass hierarchy of the $v_2$ can be observed: the heavier particle, in this case the J/$\psi$, has the smaller $v_2$, and the lighter the particle, the larger the elliptic flow. In the high $p_\textsubscript{T}$ region, this ordering disappears and the elliptic flow is independent from the particle species.

Regarding the triangular flow, a clear non-zero $v_3$ is observed for the J/$\psi$, indicating that the initial  state energy-density fluctuations reflect also in the anisotropic flow of charm quarks. The same mass hierarchy as in the case of the $v_2$ is observed at low $p_\textsubscript{T}$.

\begin{figure}[htb]
\begin{center}
\vspace{9pt}
\includegraphics[scale=0.25]{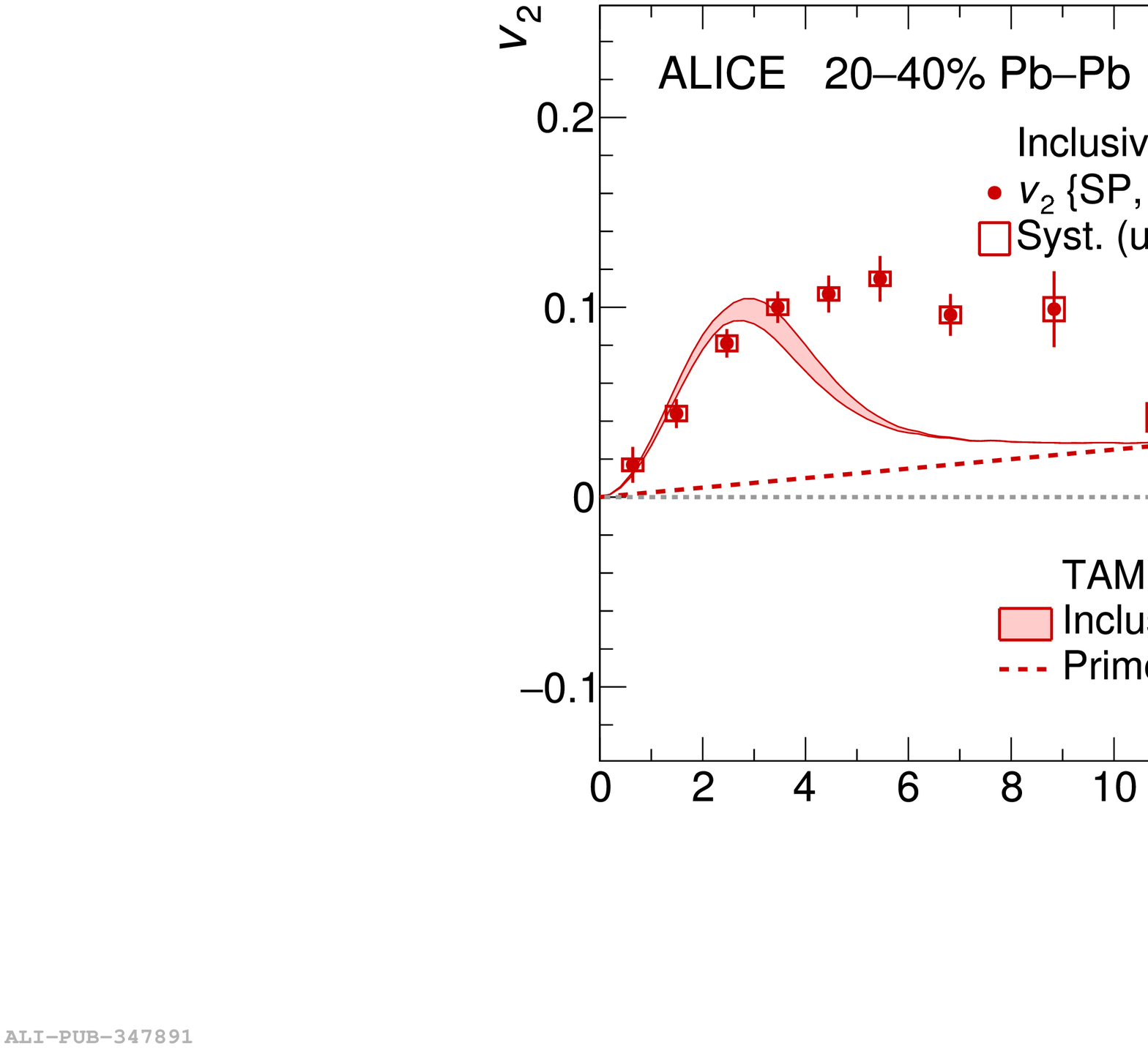}
\includegraphics[scale=0.24]{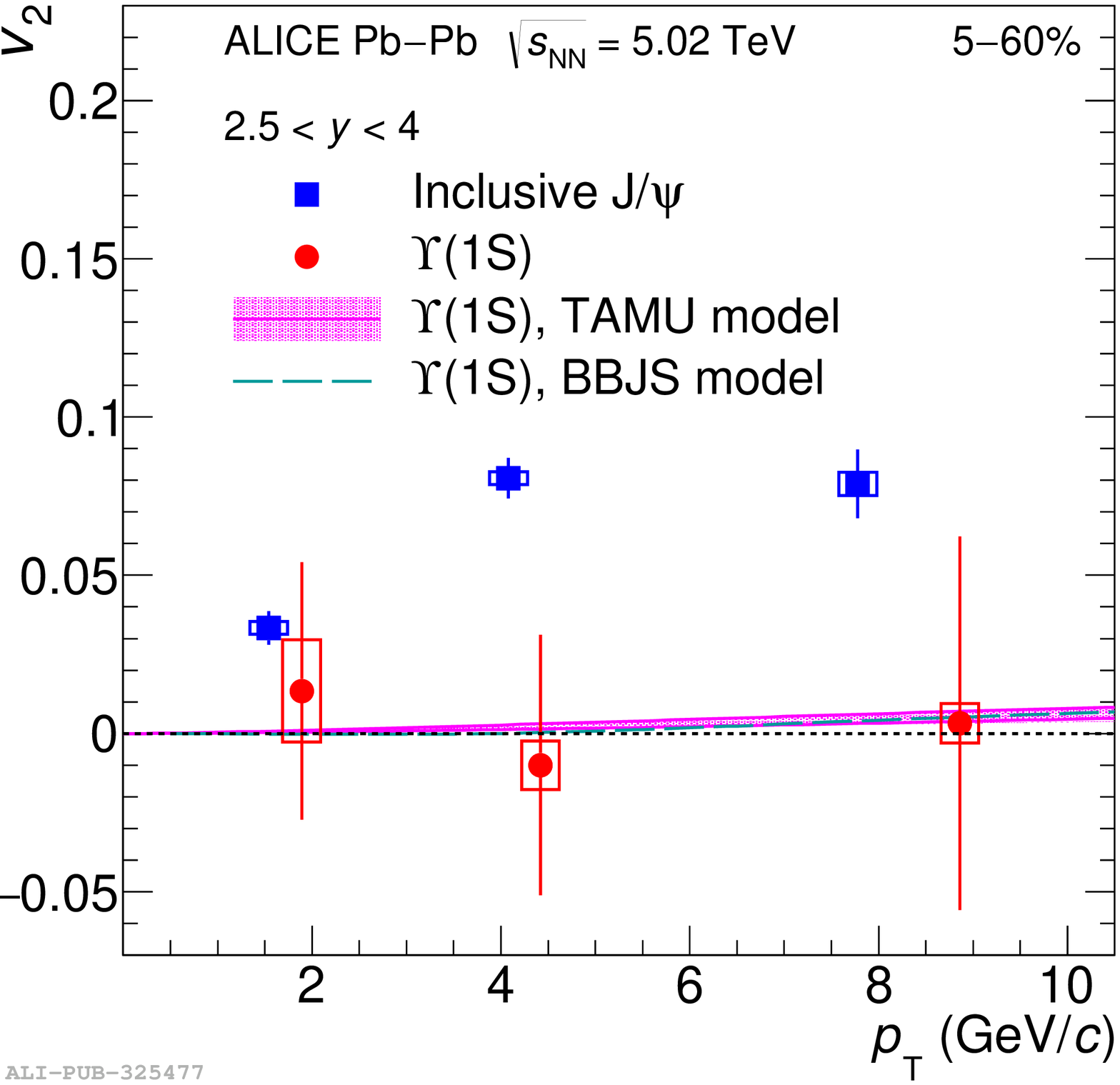}
\caption{(Top) J/$\psi$ elliptic flow as a function of $p_{\rm{T}}$~\cite{Acharya:2020jil} compared with transport model calculations~\cite{Du:2017qkv}. (Bottom) $\Upsilon$ elliptic flow as a function of  $p_{\rm{T}}$~\cite{Acharya:2019hlv} compared with the J/$\psi$ one and  model calculations~\cite{Du:2017qkv,Bhaduri:2018iwr}.}
\label{fig:V2_Models}
\end{center}
\end{figure}

The J/$\psi$ elliptic flow is compared with transport model predictions~\cite{Du:2017qkv} in Fig.~\ref{fig:V2_Models}. It appears that the model is in good agreement with the data for $p_{\rm{T}} < 4$~GeV/$c$, but strongly underestimates the data at high $p_{\rm{T}}$.

In the bottom panel of Fig.~\ref{fig:V2_Models}, the elliptic flow of the $\Upsilon$(1S) is presented as a function of $p_{\rm{T}}$, compared with the J/$\psi$ $v_2$ and transport model predictions~\cite{Du:2017qkv,Bhaduri:2018iwr}. The $\Upsilon$(1S) $v_2$ is compatible with zero, and is in agreement with the transport model predictions, which include little or no regeneration for the $\Upsilon$(1S). However, the large uncertainties on the measurement prevent from drawing any firm conclusion. Moreover, models where the b quark fully thermalize also predict a very small $v_2$ for the $\Upsilon$(1S) in the measured $p_{\rm{T}}$ region~\cite{Reygers:2019aul}. Therefore, more precise measurements are required to determine wether the beauty quark exhibits collective behaviour.

\section{Conclusion}
An overview of some of the latest results on quarkonium production in ALICE has been presented. In pp collisions at $\sqrt{s} = 13$~TeV, a new measurement of the J/$\psi$ cross section has been performed as well as a new measurement of the J/$\psi$ yield as a function of charged-particle multiplicity at mid-rapidity. It shows a faster than linear increase that is qualitatively reproduced by the different models, but the exact mechanism has yet to be understood. In p--Pb collisions at $\sqrt{s\textsubscript{NN}} = 8.16$~TeV, $R_{\rm{pPb}}$ has been measured for the J/$\psi$, $\psi$(2S), and $\Upsilon$(1S) quarkonium states. Models including final state effects can describe the two charmonium states at both forward and backwards rapidity. Finally, in Pb--Pb collisions at $\sqrt{s\textsubscript{NN}} = 5.02$~TeV, the J/$\psi$ $R_{\rm{AA}}$ has been measured at mid-rapidity. The experimental data require, beyond the effect of color screening and deconfinement, a dominant contribution from regeneration. Finally, the measurement of the J/$\psi$ elliptic and triangular flow shows a positive $v_2$ and $v_3$, and a mass hierarchy at low $p_{\rm{T}}$ in both cases. The results are in line with the regeneration scenario  except at high $p_{\rm{T}}$, where the results exhibit a large $v_2$. The $\Upsilon$(1S) $v_2$ has also been measured up to $p_{\rm{T}} = 10$~GeV/$c$ and is compatible with 0, which is expected by models.

\nocite{*}
\bibliographystyle{elsarticle-num}
\bibliography{Feuillard_V.bib}

\end{document}